\documentclass[aip,apl,amsmath,preprintnumbers,amssymb,superscriptaddress,reprint,twocolumn]{revtex4}
\usepackage{graphicx}
\usepackage{dcolumn} 
\usepackage{bm}
\usepackage[latin1]{inputenc}
\usepackage{verbatim}
\usepackage{amsmath}
\usepackage{color}

\begin{document}

\title{Second-harmonic generation as probe for structural and electronic properties of buried GaP/Si(001) interfaces}
%
\author{K.~Brixius}
\author{A.~Beyer}
\author{G.~Mette}
\author{J.~G{\"u}dde}
\affiliation{Fachbereich Physik und Zentrum f{\"u}r
Materialwissenschaften, Philipps-Universit{\"a}t, D-35032 Marburg,
Germany}
\author{M.~D{\"u}rr}
\affiliation{Institut für Angewandte Physik,
Justus-Liebig-Universit{\"a}t Gießen, D-35392 Gießen, Germany}
\author{W.~Stolz}
\author{K.~Volz}
\author{U.~H{\"o}fer}
\email[]{hoefer@physik.uni-marburg.de} \affiliation{Fachbereich
Physik und Zentrum f{\"u}r Materialwissenschaften,
Philipps-Universit{\"a}t, D-35032 Marburg, Germany}
\date{\today}
%
\begin{abstract}
Optical second-harmonic generation (SHG) is demonstrated to be a
sensitive probe of the buried interface between the
lattice-matched semiconductors gallium phosphide and silicon with
(001) orientation. {\it Ex situ} rotational anisotropy
measurements on
 GaP/Si heterostructures show a strong isotropic component of the
second-harmonic response not present for pure Si(001)or GaP(001).
 The strength of the overlaying anisotropic response directly
correlates with the quality of the interface as determined by
atomically resolved scanning transmission electron microscopy.
Systematic comparison of samples fabricated under different growth
conditions in metal-organic vapor phase epitaxy reveals that the
anisotropy for different polarization combinations can be used as
a selective fingerprint for the occurrence of anti-phase domains
and twins. This all-optical technique can be applied as an {\it in
situ} and non-invasive monitor even during growth.
\end{abstract}
\maketitle

%
%
%
%
%

The investigation of interfacial properties is one of the most
challenging tasks in modern material science and engineering of
electronic devices.
 Within the last few years large efforts have been made on the
structural analysis of interfaces by means of cross-sectional
scanning tunneling microscopy, atomic force microscopy or scanning
transmission electron microscopy (STEM).
 The latter technique is particularly suited for this purpose
because it can yield element specific information about interfaces
of inorganic materials with atomic
resolution~\cite{Voyles02nat,Krivanek10nat}.
 However, all these methods require a preceding invasive preparation of the
samples in order to access the respective interface in {\it ex situ}
experiments.

 In contrast to these methods, all-optical techniques allow a non-invasive
{\it in situ} access to interfaces, which is basically only limited
by the optical absorption length of the investigated material
system.
 Moreover, even-order nonlinear optical techniques such as second-harmonic
generation (SHG) intrinsically have a high interface sensitivity
owing to the symmetry breaking at the crystal surface and at
interfaces~\cite{Heinz91,Heinz89prl,Lupke99ssr}.
 The interface sensitivity is particularly high for centrosymmetric
materials where the bulk second-order susceptibility $\chi^{(2)}$
vanishes in dipole approximation.
 However, even in non-centrosymmetric polar materials
with strong nonlinear bulk response such as GaAs or GaP, high
interface sensitivity can be achieved by suppressing the bulk
response for specific polarization combinations of the incoming
fundamental and generated second-harmonic (SH)
light~\cite{Stehlin88ol,Yeganeh92prl}.

Here, we combine STEM and SHG to investigate the interface between
the III/V-material gallium phosphide and silicon which represents a
model system for the growth of a polar on a nonpolar semiconductor
and which promises many application for optoelectronic applications
based on silicon electronics~\cite{Liebich11apl,
Supplie15jpcl,Wang15apl,Romanyuk16prb,Kumar17prl,Lucci18prmat}.
 This system is of particular interest because the lattice mismatch between
GaP(001) and Si(001) is small.
 Therefore, strain induced defects can be neglected.
 Nevertheless, the heteroepitaxial growth is challenging because the
interface is not automatically charge neutral and anti-phase domains
(APDs) can be formed at monoatomic steps of the
substrate~\cite{Kroemer81jcg,Volz11jcg}.
 Moreover, stacking faults and twins can be formed in the GaP film.
The latter occur if a part of the crystal is rotated with respect to
the main crystal orientation.
 The structure of these defects has been extensively
investigated by means of
STEM~\cite{Beyer11jap,Beyer13apl,Beyer16cm}.
 In order to optimize the conditions for a defect-free growth,
however, an {\it in situ} technique for the evaluation of the
interface quality during growth would represent a great progress
towards an detailed understanding of the interface formation and for
the preparation of highly efficient devices.

Here, we show that SHG is able to non-invasively investigate the
properties of the buried interface between GaP and Si and that it is
possible to directly relate the SHG anisotropy to the interface
quality.
 For this purpose we have prepared GaP/Si samples by
metal-organic vapor phase epitaxy (MOVPE) under different growth
conditions that give rise to specific defects such as APDs or twins.
We present sets of the rotational anisotropy of the second-harmonic
intensity for different polarization combinations of the incoming
fundamental and generated second harmonic in reflection and compare
these results with atomically resolved STEM measurements.
 The combination of both experimental techniques allows us to
identify the occurrence of specific defects in the second-harmonic
anisotropy for particular polarization combinations.
 This makes it feasible to use second-harmonic anisotropy as an
{\it in situ}, non-invasive probe for the interface quality even
during growth. It would be complementary to other optical techniques
like reflection anisotropy spectroscopy which is particularly used
to monitor the surface properties under growth conditions in a MOVPE
reactor~\cite{Weight05rpp}.

\begin{figure}[tb]
\includegraphics[width = 0.95\columnwidth]{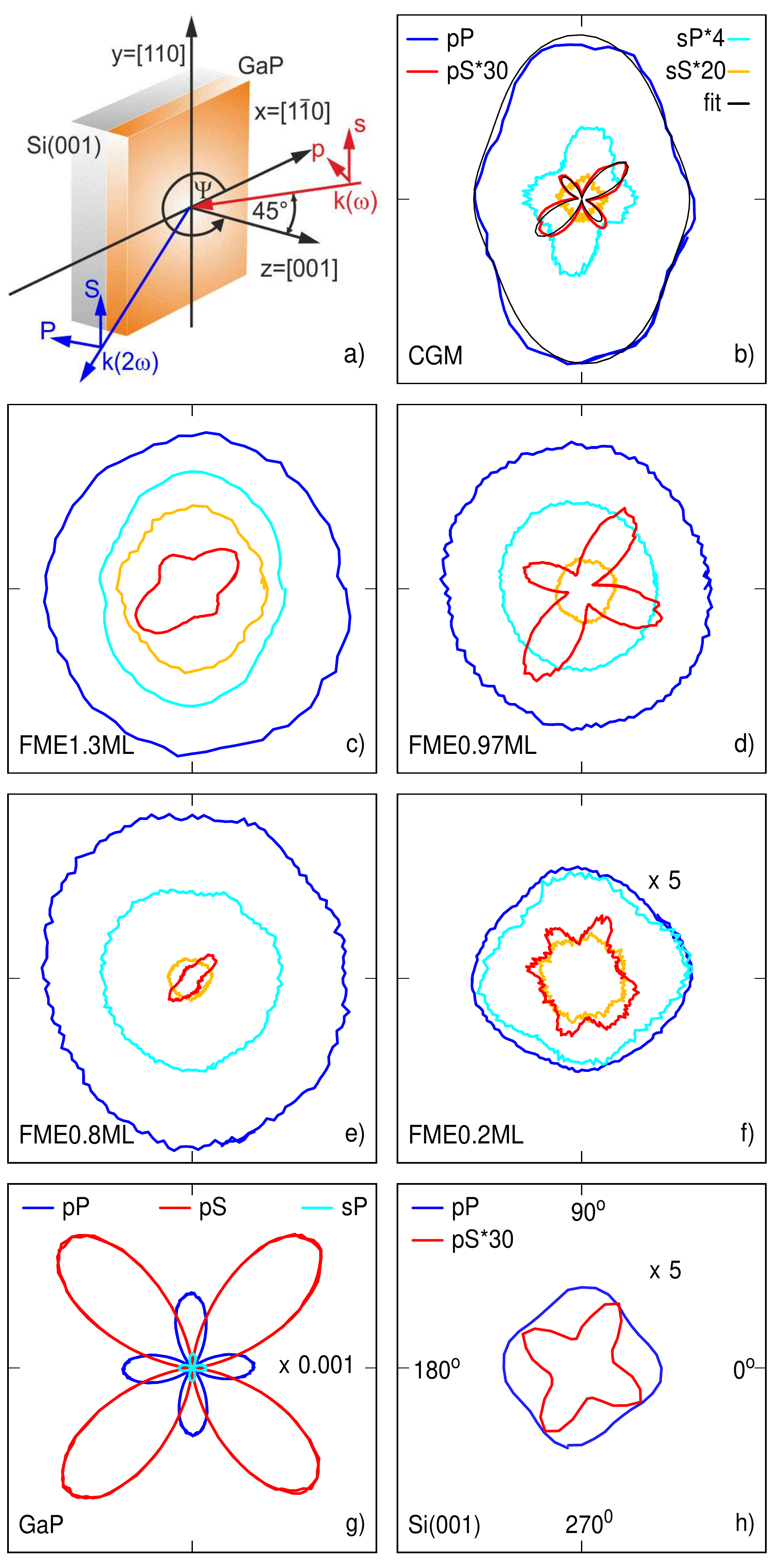}
\caption{(Color online) (a) Experimental setup.
 (b)-(f) Polar plots of the rotational SH anisotropy from five different
GaP/Si(001) samples for different polarization combinations
of the incoming fundamental and the detected SH as denoted in (b).
 For comparison, (g) and (h) show the signal of a GaP(001) wafer and a Si(001) wafer,
respectively.
  The signals of the different polarization combinations have been scaled
as denoted in the different figures for better comparison.
 Black lines in (a) exemplarily show fits of Eq.~\ref{eq_sine_cosine}
 for the polarization combinations $pP$ and $pS$.
 \label{fig5} }
\end{figure}

 The experiments were performed under ambient conditions using 50-fs
laser pulses generated by a femtosecond Ti:Sapphire laser amplifier
system operating at 800\,nm at a repetition rate of 15 kHz.
 The linear polarized laser beam at fundamental frequency $\omega$
was focussed under an angle of $45^\circ$ onto the sample as
illustrated in Fig.\ref{fig5} (a).
 The generated second-harmonic light at frequency $2\omega$ was
observed in reflection for a chosen combination of input and output
polarization, e.g. $p$-polarized incident light and $s$-polarized
$2\omega$-light (abbreviated by $pS$).
 Standard boxcar-integrator technique was used for detection
as described in detail in Ref.~\onlinecite{Stepan05ss}.
 The SH signal has been normalized with respect to a reference
that was generated in a quartz crystal in $pP$ configuration.
 The rotational anisotropy of the SH intensity was measured by rotating
the sample around the surface normal ($z$-axis) as characterized by
the azimuthal angle $\Psi$.
 The samples were oriented such that for $\Psi=0^\circ$ ($180^\circ$)
the $[\bar110]$- ($[1\bar10]$)-direction was lying within the plane
of incidence (cf. Fig.\ref{fig5} (a)).
 The fluence of the incident laser radiation was kept at least one
order of magnitude below $100~\rm{mJ/cm^2}$ which we determined as
the threshold for multishot damage of our samples.

The investigated GaP/Si heterostructures were grown by metal-organic
vapor phase epitaxy (MOVPE) in an Aixtron AIX 200 GFR reactor using
triethylgallium (TEGa) and tertiary-butyl phosphine (TBP) as
precursors for gallium and phosphorus, respectively.
 Four samples were prepared by using flow-rate modulated epitaxy
(FME), for which TEGa and the TBP were supplied intermittently in a
total of sixteen cycles, starting with TEGa.
 The supply of GaP per cycle was varied between $0.2$ to $1.36$ monolayer
by a variation of the partial pressure of TEGa between
$7.6\times10^{-4}$ and $7.6\times10^{-3}~\rm{mbar}$ while holding
the TBP pressure constant at $0.91~\rm{mbar}$.
 In the following, these samples are referred to as FME0.20ML and so forth.
 For one sample, the continuous growth mode (CGM) was used,
in which TEGa and TBP were applied simultaneously.
 The total thickness of the layers varied between 0.8 and 4.5~nm
as determined by X-ray reflection in an X´Pert Pro diffractometer.
The structural characterization of the GaP/Si interfaces was carried
out in a JEOL 2200FS scanning transmission electron microscope
operating at $200~\rm{kV}$.
 This STEM can achieve atomic resolution in the high-angle
annular dark-field imaging mode by aberration corrections of the
probe forming lenses. For comparison, we also studied the SH
response of the individual materials.  In case of Si, a native oxide
covered Si(001) wafer was used. In case of GaP, a $1~\rm{\mu m}$
thick GaP film was grown epitaxially on a GaP(001) wafer.


The azimuthal dependence of the SH intensity for all samples and
polarization combinations is presented in Fig.~\ref{fig5} (b)-(h).
 These rotation patterns offer a direct access to the symmetry
and rotational anisotropy of the second-order nonlinear
response~\cite{Heinz91}.
 For the GaP wafer (Fig.~\ref{fig5} (g)), a four-fold anisotropy was found for the
pola\-ri\-zation combinations $pP$, $pS$ and $sP$ in consistence
with other reports~\cite{Stehlin88ol}.
 This anisotropy results from the $\bar43m$-symmetry group of
zincblende crystals and is a pure bulk dipole contribution.
 The $pS$-combination shows the strongest SHG;
 it vanishes completely if the polarization of the fundamental
is along the $[110]$- and $[\bar110]$-direction. In contrast to the
GaP wafer, the SH response of the native oxide covered Si(001) wafer
(Fig.~\ref{fig5} (h)) is weaker by about four orders of magnitude.
 Its fourfold anisotropy is known to originate from the bulk
quadrupole contribution~\cite{Tom83prl} whereas the isotropic
contribution is attributed to the SiO$_2$ layer or the Si/SiO$_2$
interface~\cite{Lupke99ssr}.
 Within our detection limit, no SH signal could be observed in case of
the $sS$-combination for both the GaP and the Si wafer.

Compared to the two individual wafers, the SH response of the five
he\-te\-ro\-epi\-taxially grown films of GaP on Si(001)
(Fig.~\ref{fig5} (b)-(f)) differs remarkably, both in overall
intensity and in particular in anisotropy.
 Most prominent is the dominant isotropic contribution in
the patterns for the $pP$ (blue lines) and the $sP$ (cyan lines)
polarization combination.
 Except for the thinnest sample FME0.2ML, this signal is
considerably larger than the response of the Si wafer.
 Moreover, it by far outreaches the almost invisible four-fold
anisotropic bulk contribution of the GaP films on Si. We therefore
assign the isotropic signal to the interface between GaP and Si.
Since the GaP wafer was grown under similar conditions as the GaP/Si
samples but does not show a measurable isotropic signal, we exclude
that the GaP surface contributes significantly to the observed
isotropic SH component. This interpretation is supported by
additional measurements for GaP layers with thicknesses of up to 65
nm (not shown) which showed that the isotropic contribution of the
SH response decreases exponentially with the GaP layer thickness.

A second-order nonlinear response of the interface between two
semiconductors can arise either from the presence of a static
electric field due to the band alignment of both
materials~\cite{Ishioka16apl} or from the modification of the
chemical bonds at the interface. For the GaP/Si interface, the
latter is related to the hybridized sp$^3$ orbitals that are
distorted due to the Ga-Si and P-Si bonds.
 Similar, although quantitatively stronger effects, have been observed for
the clean Si surface, where a comparable symmetry break due to
surface reconstruction results in an increase of the SH intensity by
a factor of $\sim100$~\cite{Hofer96apa}.
 In our case of the GaP/Si interface, the increase compared to the
oxidized Si surface is only a factor of $\sim10$.
 This weaker effect might be connected to a cancelation due to
bond distortions in opposite directions.
 An electric field at the interface gives rise to the
EFISH-effect~\cite{Lee67prl,Aktsip96}, which also predominantly
contributes to the p-polarized component of the SH intensity because
it particularly affects the nonlinear response perpendicular to the
interface. It has been demonstrated that fields due to depletion
layers can exceed the nonlinear bulk response, even in case of
strong bulk signals for III-V materials like
GaAs~\cite{Germer97prb,Aktsip99prb}.
Most probably, both distorted bonds as well as electric fields at
the interface contribute to the observed isotropic response.

 The $pS$ contribution (red lines in Fig.\ref{fig5}) of the SH
response seems to be less affected by the interface because its
shape is similar to the response of the Si and the GaP wafer.
 Its strength, however, is almost five orders of magnitude
smaller as compared to the bulk response of the GaP wafer.
 This is surprising because the penetration depth of the fundamental
light $\delta_{\rm{GaP,800nm}}$ is about $160~\rm{\mu m}$ and the SH
intensity increases up to a film thickness of $\sim 30~\rm{nm}$
before a reduction due to a phase shift between the $\omega$- and
$2\omega$-light sets in.
 Therefore, the GaP bulk contribution from thin films of a few nm
thickness should only be smaller by about one order of magnitude
compared to the thick wafer.
 This strong reduction of the bulk response is most likely being
caused by a destructive interference due to the appearance of
anti-phase domains because the phase of the SH response originating
from P-polar GaP is shifted by about $\pi$ in comparison to the
response of the Ga-polar GaP. This makes SHG very sensitive to APDs
as it was demonstrated for thick GaAs films grown on
Si(001)~\cite{Lei13apl}.

 Beyond these general differences between the SH response of the
individual wafers and the GaP/Si samples, the different
heterostructures show characteristic variations for all measured
polarization combinations.
 In the following we show that these individual differences of the SH
response can be related to structural differences that have been
obtained by STEM.
 This makes it possible to use the nonlinear response as a
fingerprint of the GaP/Si interfaces.

 For this purpose a symmetry analysis has been applied that
quantifies the different isotropic and anisotropic contributions to
the nonlinear response. Phenomenologically, the second-harmonic
intensity  $I_{ij}(2\omega)$ in reflection for a given polarization
combination $ij$ can be written as a function of the incidence
intensity $I_i(\omega)$ as $I_{ij}(2\omega)\propto |\chi_{{\rm
eff},ij}^{(2)}|^2 I_i^2(\omega)$ where $\chi_{{\rm eff},ij}^{(2)}$
is an effective second-order nonlinear susceptibility tensor of
third rank that includes all contributions (surface, interface,
bulk, EFISH, etc.) to the second-order nonlinear
polarization~\cite{Heinz91} even if the intrinsic material response
of the bulk and EFISH contributions might be in the most general
case described by higher-order tensors of the nonlinear
susceptibility~\cite{Sipe87prb,Aktsip96}.
 The dependence of $I_{ij}(2\omega)$ on the azimuthal angle $\Psi$
can then be written as a Fourier expansion up to the fourth order
\begin{equation}
I_{\rm{ij}}(2\omega)\propto \Big\vert a_{\rm{ij}} + \sum_{m=1}^4  b_{\rm{ij}}^{(m)} \cos{m\Psi} + c_{\rm{ij}}^{(m)} \sin{m\Psi}\Big|^2,
\label{eq_sine_cosine}
\end{equation}
where the isotropic ($a_{\rm{ij}}$) and anisotropic coefficients
($b_{\rm{ij}}^{(m)}$, $c_{\rm{ij}}^{(m)}$) contain all optical
properties, in particular the corresponding components of the
nonlinear susceptibility tensors as well as Fresnel coefficients.
 For (001) oriented samples, odd orders $m$ can be excluded.

Table~\ref{tab_coefficients} lists the expected non-vanishing
isotropic and anisotropic coefficients at the different polarization
combinations for all considered SHG sources, i.e. Si-surface, -bulk
and -EFISH as well as GaP-bulk and -EFISH.

\begin{table}[b]
\begin{center}
     \begin{tabular*}{\columnwidth}[t]{
    @{\extracolsep\fill}l
    @{\extracolsep\fill}c
    @{\extracolsep\fill}c
    @{\extracolsep\fill}c
    @{\extracolsep\fill}c
}

\hline\hline origin of SHG & $pP$ & $pS$ & $sP$ & $sS$
 \\

\hline\hline

Si surface (single-domain) & $a_{pP}^{},c_{pP}^{(2)}$ &
$b_{pS}^{(2)}$ & $a_{sP}^{},c_{sP}^{(2)}$ & $-$ \cr

Si surface (multi-domain) & $a_{pP}^{}$ & $-$ & $a_{sP}^{}$ & $-$
\cr

Si bulk (quadrupole) &$a_{pP}^{},c_{pP}^{(4)}$ & $b_{pS}^{(4)}$ &
$a_{sP}^{},c_{sP}^{(4)}$ & $b_{sS}^{(4)}$ \cr

Si EFISH $\chi^{(3)}$ & $a_{pP}^{}$ & $-$ &$a_{sP}^{}$ &$-$ \cr

\hline

GaP bulk & $b_{pP}^{(2)}$ & $c_{pS}^{(2)}$ & $b_{sP}^{(2)}$ & $-$
\cr

GaP bulk (quadrupole)  & $a_{pP}^{},c_{pP}^{(4)}$& $b_{pS}^{(4)}$ &
$a_{sP}^{},c_{sP}^{(4)}$ & $b_{sS}^{(4)}$\cr

GaP EFISH $\chi^{(3)}$ & $a_{pP}^{}$ & $-$ &$a_{sP}^{}$ &$-$ 

 \cr\hline\hline
\end{tabular*}
 \begin{minipage}{\columnwidth}
\caption[]{\label{tab_coefficients} \noindent Isotropic and
anisotropic contributions to the SHG from surface, bulk and EFISH of
Si(001) and GaP(001) for the different polarization combinations.}
    \end{minipage}
  \end{center}
\end{table}

 Within this model the azimuthal dependence of the SH intensity can be well
described for all polarization combinations by a least-square fit of
the coefficients $a_{ij}$, $c_{ij}^{(2)}$, $b_{ij}^{(2)}$ and
$b_{ij}^{(4)}$, i.e.\ omitting the $c_{ij}^{(4)}$. Two examples of
these fits for the polarization combinations $pP$ and $pS$ are
shown in Fig.~\ref{fig5}~(b) (black lines).

\begin{figure}[t]
\includegraphics[width = 0.95\columnwidth]{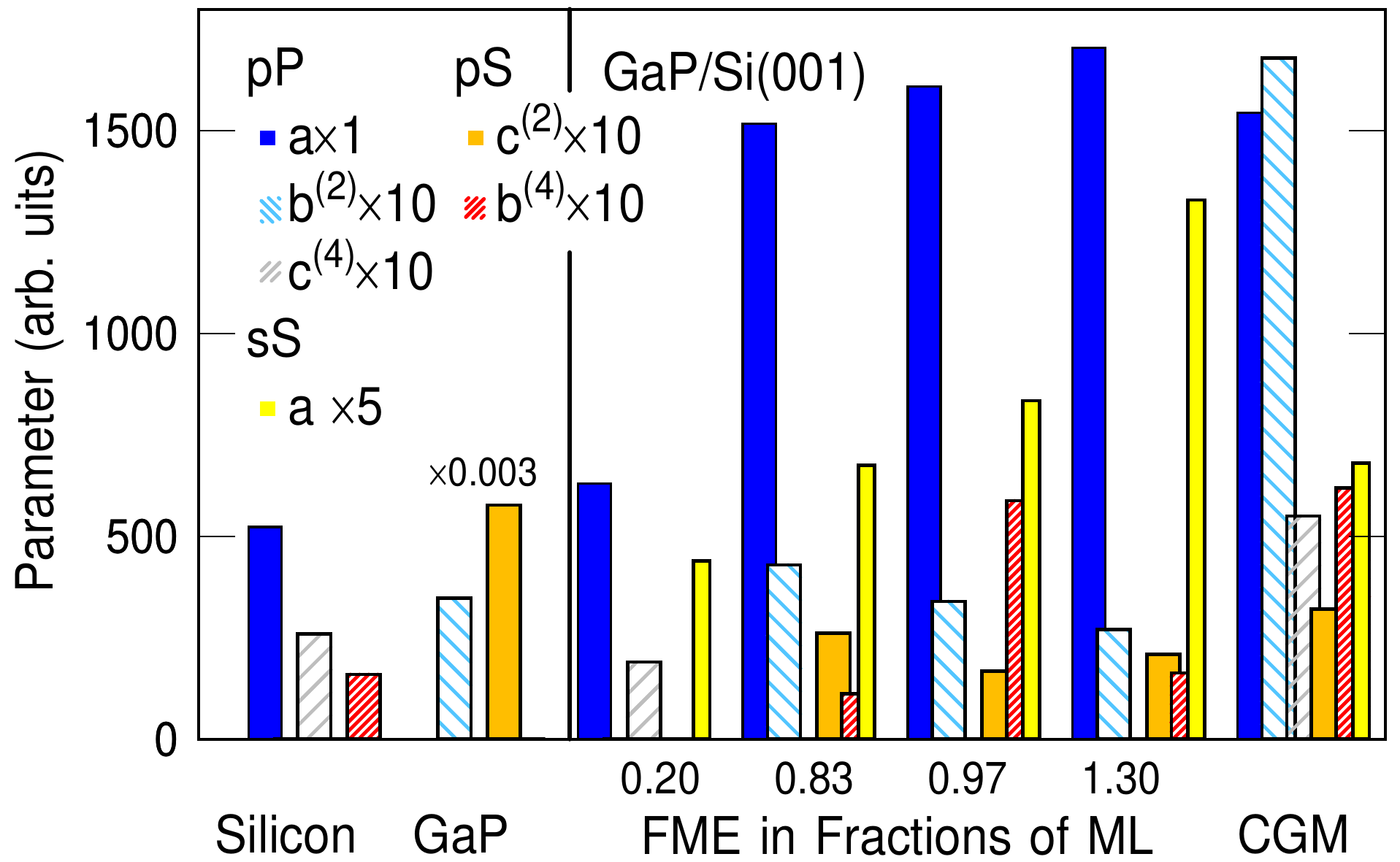}
\caption{(Color online) Isotropic and anisotropic coefficients for all
investigated samples as obtained by a least-square fit of Eq.~\ref{eq_sine_cosine} for the
polarization combinations $pP$, $pS$, and $sS$.
 \label{barplot}}
\end{figure}

 Figure~\ref{barplot} summarizes the fitting results for the
polarization combinations $pP$, $pS$, and $sS$.
 For all heterostructures, the interface specific isotropic parameter $a_{\rm{pP}}$
(blue bars) dominates the total response.
 For the sample with the smallest GaP thickness of 0.8~nm (FME0.20ML), the
magnitude of this contribution is comparable to that of the Si
wafer.
 For the other GaP/Si samples including the CGM sample,
$a_{\rm{pP}}$ shows only a small variation.
 This suggests that the interface properties do not change
substantially for GaP layer thicknesses of $1.6$ to $4.5~\rm{nm}$.

\begin{figure}[!b]
\includegraphics[width = 0.95\columnwidth]{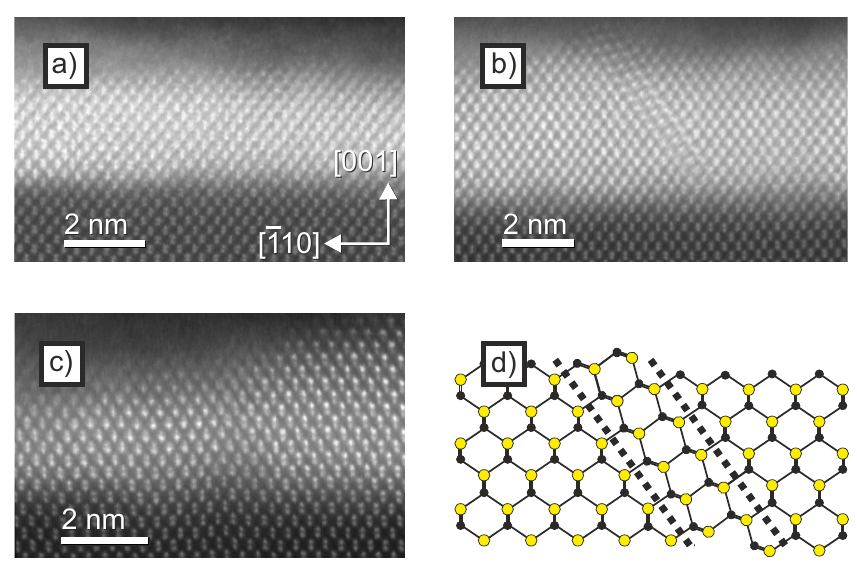}
\caption{(Color online) STEM images of the GaP/Si interface for the
samples (a) FME0.97ML, (b) CGM, and (c) FME1.30ML. (d) Scheme of
twin defect in GaP. Ga and P atoms are depicted by yellow and black
dots, respectively.\label{ga-drops}}
\end{figure}

The STEM images show, however, that the number of defects in the
interface-near region of the FEM samples increases considerably for
a GaP supply of more than 1~ML per cycle. This is shown exemplary in
Fig.~\ref{ga-drops} (a) and (c) by comparing the STEM images of
FEM0.97ML and FEM1.30ML. Whereas the GaP film of sample FEM0.97ML
shows no visible defects and a rather abrupt interface, two APDs of
opposite polarity can be identified in the left and right part of
the STEM image of sample FEM1.30ML.
 These APDs can form if an excess of TEGa etches Ga droplets
into the Si crystal which then serve as nuclei for
APDs~\cite{Werner14jcg}.
 The formation of APDs causes an in-plane symmetry reduction
that affects predominantly the in-plane ($s$-polarized) SHG
components.
 This is directly reflected by the strong increase of the isotropic
$a_{\rm{sS}}$ contribution (yellow bars in Fig.~\ref{barplot}) with
increasing TEGa pressure which is characterized by the supply of GaP
monolayer per cycle.
 If APDs grow straight up through the crystal, they typically have
a rectangular shape with similar edge lengths in the $[110]$- and
$[\bar110]$-directions~\cite{Werner14jcg}.
 Such fourfold anisotropy is in fact slightly visible in the $sS$-component
of the rotational SH anisotropy for sample FME1.30ML (yellow line in
Fig.~\ref{fig5}~(c)).

The STEM measurements on the CGM sample exhibit much less APDs and,
consequently, only a small magnitude of $a_{sS}$ is observed in SHG.
 In this growth mode, however, the formation of twins is favored
as discussed in detail in Ref~.\onlinecite{Beyer11jap}.
 Such a twin, which is sketched in Fig.~\ref{ga-drops} (d), can be
clearly identified in an STEM image of the CGM sample shown in
Fig.~\ref{ga-drops} (b).
 It gives rise to a two-fold anisotropy between the $[\bar110]$ and
$[110]$ directions because twins predominately form on Ga-terminated
$\{111\}$A planes~\cite{Beyer11jap,Skibit12jap}.
 This anisotropy is characterized by a large magnitude of $b^{(2)}_{\rm{pP}}$
(cyan/white bars in Fig.~\ref{barplot}) as observed for the CGM
sample. In contrast, $b^{(2)}_{\rm{pP}}$ is comparatively small for
all FME samples.

 The anisotropy for the $pS$ polarization combination is
characterized by the parameters $c^{(2)}_{\rm{pS}}$ and
$b^{(4)}_{\rm{pS}}$ which are depicted by orange and red/white bars
in Fig.~\ref{barplot}, respectively.
 These parameters might be further indicators for the film quality.
They show, however, no clear trend for the different growth modes
and cannot be correlated to specific defects observed in STEM.

 In summary, we have shown that optical second-harmonic generation can be
successfully used as a sensitive, non-invasive probe for the quality
of the interface between thin films of GaP and Si(001).
 The rotational anisotropy of the second-harmonic intensity can be
correlated to specific defects that form at different growth
conditions in metal-organic vapor phase epitaxy.
 In particular, the occurrence of anti-phase domains and twins
are clearly identified. This all-optical technique can be applied as
an {\it in situ} monitor of the growth process complementary to
reflection anisotropy spectroscopy.

%
We gratefully acknowledge funding by the Deutsche
Forschungsgemeinschaft through the SFB 1083.
%
\bibliographystyle{prsty}


\end{document}